\documentstyle[12pt]{article}
\newcommand{\be}{\begin{eqnarray}}
\newcommand{\ee}{\end{eqnarray}}
\newcommand{\half}{\frac{1}{2}}

\newcommand{\at}{\tilde a}
\newcommand{\Qt}{\tilde Q}

\newcommand{\phit}{\tilde {\phi}}
\newcommand{\chit}{\tilde {\chi}}
\newcommand{\phib}{\bar {\phi}}

\newcommand{\vp}{\varphi}
\newcommand{\zb}{\bar z}
\newcommand{\p}{\partial}
\newcommand{\e}{\varepsilon}
\newcommand{\texthalf}{\textstyle{1 \over 2}}

\begin{document}
\gdef\journal#1, #2, #3, 1#4#5#6{{#1} {\bf #2}, #3 (1#4#5#6)}
\gdef\ibid#1, #2, 1#3#4#5{{#1} (1#3#4#5) #2}

\begin{flushright}  
Oslo-SHS-96-5 \\
hep-th/9605157\\
\end{flushright}


\vskip 5mm
\centerline{\large\bf Model of statistically coupled chiral fields on the 
circle} 
\vskip 10mm
\centerline{ {\large Serguei B. Isakov$^{\star 1}$ 
                 and Susanne Viefers$^{\dagger}$ }}
\vskip 3mm
\centerline{{\em Centre for Advanced Study, Drammensveien 78, 0271 Oslo, 
Norway}}
{\em \centerline{and}  
\centerline{
$^{\star}$Medical Radiological Research Center, Obninsk, Kaluga Region
           249020, Russia}
\centerline{
$^{\dagger}$Institute of Physics,  University of Oslo,
P.O. Box 1048 Blindern, N-0316 Oslo,
Norway}}

\vskip 15mm
\newcommand \sss {\mbox{ $<\overline{s}s>$} }
\def\fk{\mbox{ $f_K$} }
\centerline{ ABSTRACT}
\vskip 3mm
Starting from a  field theoretical description of multicomponent anyons with 
mutual 
statistical 
interactions in the lowest Landau level, we construct a model of interacting 
chiral fields on the circle, with the energy spectrum characterized by a 
symmetric matrix $g_{\alpha\beta}$ with nonnegative entries. Being represented 
in a free form, the model provides a field theoretical realization of (ideal) 
fractional exclusion statistics for particles with linear dispersion, with 
$g_{\alpha\beta}$ as a statistics matrix.  
We derive the bosonized form of the model and discuss the relation   
to the effective low-energy description of the edge excitations 
for  abelian  fractional quantum Hall states in multilayer systems.

\vfil
\noindent
$^{\dagger}$Supported by the Norwegian Research Council. 

\noindent
$^1$Address from 1 August 1996: Institute of Physics,  University of Oslo,
P.O. Box 1048 Blindern, N-0316 Oslo, Norway.

\newpage


\noindent
{\bf 1. Introduction}
\medskip

Recently a field theory of anyons confined to the 
lowest Landau level (LLL) of  a strong magnetic field was formulated 
\cite{HLV}. 
The authors of Ref.~\cite{HLV} next mapped  their field theory (which is 
effectively one dimensional) onto a circle  obtaining  a 
theory of a chiral fermion field with linear dispersion.  
The problem  involves  a  (statistics) parameter 
$g$ entering the wave function of anyons in the LLL
$\Psi=\prod_{i<j}(z_i-z_j)^{g}\Phi_+(\{z_i\})e^{-\frac12\sum_i|z_i^2|}$, 
where $\Phi_+(\{z_i\})$ is a 
symmetric function of its  arguments. The requirement that the function 
$\Phi_+$ should be a polynomial for all the states in the LLL implies  
that $g\in [0,2)$. 
Remarkably, the field theory on the circle remains consistent in a wider 
interval of range of the parameter $g$, for all $g >0$. 
The motivation to consider such an extension of the theory is the  
similarity  of the  above wave function with $g$ odd to  
Laughlin wave functions of fractional quantum Hall (FQH) states 
\cite{Laughlin}. 
It suggests that for $g$ odd, the model of a chiral field discussed in 
\cite{HLV} 
may be used  to  describe edge excitations for Laughlin states.   
Indeed, a bosonization of the theory 
enables one to recover  correlation exponents as well as the low temperature 
  thermodynamics \cite{HLV} coinciding with  those obtained with the use of 
the 
(chiral) Luttinger liquid description of edge excitations \cite{Wen}
(for a recent review, see \cite{Wen-review}).  

In this paper we generalize  the model in \cite{HLV} to the case of several 
species 
of particles. In Section 2, we discuss a field theoretical description of 
anyons of several species in the LLL 
described by the  generic statistics matrix $g_{\alpha\beta}$
(with off-diagonal elements corresponding to mutual 
statistical interactions \cite{IMO}), and then map the system onto a circle. 
Upon  extending the possible values of the statistics parameters, we arrive 
at  
a model of interacting chiral fields,  with the energy spectrum     
characterized by a symmetric matrix $g_{\alpha\beta}$ with non-negative 
entries. 
As in the case of a single species, the theory 
may be reformulated in a free form 
(in terms of field operators obeying more complicated commutation relations). 
We discuss the construction of the corresponding Fock space.  
In Section 3 we show that the
structure which is  behind the above  free field formulation  is an ideal  
fractional exclusion statistics 
\cite{Haldane-ES,I-IJMPA,I-MPLB,Wu,dVO,MSh,FK-PRB}: in the thermodynamic 
limit  
the model is 
equivalent to a system of free chiral particles with linear dispersion, 
characterized by the (bosonic) statistics matrix $g_{\alpha\beta}$.

It was observed in Ref.~\cite{HLM} that 
as a result of the dimensional reduction, the system of two anyons in the LLL 
admits a description in terms of fractional statistics in one dimension. 
The latter statistics was originally introduced by Leinaas and Myrheim in the 
Heisenberg quantization of identical particles \cite{LM-HeisQuant}. It was 
also 
argued in Ref.~\cite{LM-HeisQuant} that the  system of noninteracting 
particles 
obeying fractional statistics in  one dimension    
is equivalent to a system of particles with long range inverse square 
interaction (see also \cite{Poly-NPB89}). Using the asymptotic Bethe ansatz 
for the latter model \cite{Sutherland}, an equation for the single state 
distribution function for fractional statistics in one  dimension was derived 
(note that in spite of being derived from the one-dimensional model,     
the resulting single state statistical distribution does not  make any 
reference 
to the space dimension) \cite{I-IJMPA}. The same distribution function was 
also 
derived \cite{I-MPLB,Wu} from the generalized exclusion principle 
\cite{Haldane-ES} and the statistics is now referred to as  fractional 
exclusion 
statistics. The system of anyons in the LLL is a realization of this 
statistics, as can be seen from the equation of state and  thermodynamic 
quantities 
\cite{Wu,dVO,IMO}.

The above  explains the appearance of exclusion statistics in the model 
of chiral fields on the circle which has the same quantum numbers as the  
system 
of anyons in the LLL.  Using the exclusion statistics representation of the 
model, 
we calculate the low temperature heat capacity.  
 The connection with exclusion statistics enables one to describe the 
model in terms of equations having  the form of   
asymptotic  Bethe ansatz equations \cite{Sutherland}. This implies a  
 simple  dynamics  encoded in two-body scattering phases of a step-wise form.
   
We discuss some peculiarities in calculating quantities like the dressed 
charge matrix due to a step-wise character of the two-body scattering phases.

In Section 4 we consider  the relation to the effective low-energy description
 of 
edge excitations in abelian multilayer FQH states  \cite{WenZee,Wen-review}. 
In the  bosonized form, our  model describes  a set of chiral boson fields all
propagating in the same direction with the same velocity. 
For the case of odd  diagonal entries and integer off-diagonal entries,  
the statistics matrix $g_{\alpha\beta}$ can then be identified with the    
topological matrix $K_{\alpha\beta}$ describing the edge excitations.  
This in particular implies an interpretation of edge excitations in terms of 
exclusion statistics and is consistent with the results of Fukui and Kawakami  
 who achieved the same identification between 
 $g_{\alpha\beta}$ and $K_{\alpha\beta}$
for edge excitations in hierarchical FQH states \cite{FK-PRB}.

\bigskip


\noindent
{\bf 2. Multicomponent field theory on the circle}
\medskip

Consider a system of $M$ species of anyons in the LLL, 
described by the generic statistics matrix 
$g_{\alpha\beta}$ ($\alpha, \beta=1,\dots M$), with 
$N_{\alpha}$ the  number of particles of species $\alpha$. 
The wave function of the  system has the form \cite{IMO} 
\be
\Psi = \prod_{\alpha}\prod_{i<j} 
(z_i^{\alpha}-z_j^{\alpha})^{g_{\alpha\alpha}}
       \prod_{\alpha < \beta}\prod_{i,j} 
(z_i^{\alpha}-z_j^{\beta})^{g_{\alpha\beta}}
       \Phi_+(\{z_i^{\alpha}\},\{z_i^{\beta}\}, ...)
e^{-\frac12 \sum_{\alpha}\sum_{i}{|z_i^{\alpha}|}^2} \,. 
\label{wf}\ee
Here $z_{i}^{\alpha}=\sqrt{eB/2}(x_{i}^{\alpha}+i y_{i}^{\alpha})$ is a 
dimensionless complex coordinate of the $i^{\rm th}$ particle of species 
$\alpha$ (the charges and masses of all the species of particles are assumed 
to be the same). The symmetric matrix   $g_{\alpha\beta}$ is  responsible for 
the  statistics of anyons, with non-diagonal elements corresponding to 
mutual statistical interactions between particles of distinct species.  
 The function $\Phi_+$ is 
single-valued and symmetric with respect to an  interchange of coordinates 
of the {same} species, arbitrary when
coordinates of {distinct} species are interchanged (bosonic representation). 

The wave function (\ref{wf}) acquires a phase 
$\exp(i\pi g_{\alpha\alpha})$ under an (anticlockwise) 
interchange of two particles of 
species $\alpha$ and a phase $\exp(i2\pi g_{\alpha\beta})$ under winding 
a particle of species $\alpha$ around a particle of species $\beta$ for 
$\alpha\neq\beta$ (provided 
that each of the above closed paths encloses no other particles).  
In order for the function $\Phi_+$ 
to be a polynomial for all the states in the LLL, 
one should choose the intervals $g_{\alpha\alpha}\in [0,2)$ and   
$g_{\alpha\beta}\in [0,1)$ for $\alpha\neq\beta$. 
$g_{\alpha\beta}=0$ and $g_{\alpha\beta}=\delta_{\alpha\beta}$ correspond to 
sets of  bosons and fermions, respectively.

It will be convenient for us to work in the 
fermionic representation. The latter  is obtained by 
the change 
$g_{\alpha\beta}=\delta_{\alpha\beta}+ \lambda_{\alpha\beta}$ and 
absorbing 
a Slater determinant corresponding to the $\delta_{\alpha\beta}$-part of  
$g_{\alpha\beta}$ in (\ref{wf}) 
in the definition of the function $\Phi$. 
This  converts $\Phi_+$ into the function 
$\Phi_-(\{z_i^{\alpha}\},\{z_i^{\beta}\}, ...)$ which is antisymmetric with 
respect to an interchange of coordinates within the same species of particles. 

The angular momentum for the states (\ref{wf}) has eigenvalues  
\be
L = \sum_{\alpha} \left[ \sum_{j=1}^{N_{\alpha}} n^j_{\alpha } + 
\half\lambda_{\alpha\alpha}N_{\alpha}(N_{\alpha}-1) 
+ \half\sum_{\beta\neq\alpha} \lambda_{\alpha\beta} N_{\alpha} N_{\beta} 
\right], 
\label{L(N)}\ee
where $n^j_{\alpha}$ are nonnegative integers, distinct  within each species, 
corresponding  to the  fermionic 
part of the angular momentum associated with the function $\Phi_-$.

Following Ref.~\cite{HLV}, we introduce the  fermionic field operators 
$\vp_{\alpha}(z)$ obeying the following anticommutation relations,
\be
\{\vp_{\alpha}(z),\vp_{\beta }(z')\} = 0\,,\quad
\{\vp_{\alpha}(z'),\vp_{\beta}^{\dagger}(\zb)\} = \delta_{\alpha\beta} 
\frac{1}{\pi} e^{z'\zb} \,.
\label{CR-varphi}\ee
On  expanding $\vp_{\alpha}(z)$ in powers of the angular momentum eigenstates 
\be
\vp_{\alpha}(z) = \sum_{l=0}^{\infty} \frac{1}{\sqrt{\pi l!}}\, 
\vp^{\alpha}_l z^l\,,
\label{phi-expan}\ee
these imply the usual fermionic anticommutation relations for the components: 
\be
\{ \vp^{\alpha}_l,  \vp^{\beta}_{l'}\}=
\{ \vp^{\alpha\dagger}_l,  \vp^{\beta\dagger}_{l'}\}=0\,, \quad 
\{ \vp^{\alpha}_l,  \vp^{\beta\dagger}_{l'}\}=
\delta_{\alpha\beta}\delta_{ll'}\,.
\label{Fermi}\ee 
In terms of  the field operators (\ref{CR-varphi}), 
the angular momentum can be written as 
\be
\hat L = \sum_{\alpha}\int d^2z e^{-\zb z} \vp_{\alpha}^{\dagger}(\zb)
    \left[ z\p_z + \half\sum_{\beta} \lambda_{\alpha\beta} \hat N_{\beta} 
    \right] 
\vp_{\alpha}(z)\,,
\label{L-phi}\ee
with the representation of the particle number operators as 
\be
\hat N_{\alpha} = \int d^2z e^{-\zb z} \vp_{\alpha}^{\dagger}(\zb) 
\vp_{\alpha}(z)
\label{N-phi}\ee
($d^2z\equiv dxdy$). 
By using  the expansion 
(\ref{phi-expan}) and the explicit construction of the fermionic Fock space 
generated by the operators (\ref{Fermi}), one can verify that 
the formula (\ref{L-phi}) recovers the correct angular momentum eigenvalues
(\ref{L(N)}) for  states with fixed particle numbers.  

Upon adding a harmonic potential (of frequency $\omega$), the Hamiltonian 
becomes
\be
{\cal H} = \half \omega_c^{\rm eff} \sum_{\alpha}\hat N_{\alpha} + a\hat L \,, 
\label{H}\ee
where $\omega_c^{\rm eff} = \sqrt{\omega_c^2 + 4\omega^2}$, 
$\omega_c = eB/m$ is the cyclotron frequency, 
and $a=\omega_c^{\rm eff}-\omega_c$. To leading order in 
$\omega^2/\omega_c^2$,  
which is assumed hereafter, 
\be
\omega_c^{\rm eff}\simeq\omega_c\,, \quad  a\simeq\omega^2/\omega_c\,, 
\quad\quad 
\omega \ll \omega_c \,.
\ee

The wave functions  (\ref{wf}) for $g_{\alpha\alpha}$ odd integers and 
$g_{\alpha\beta}$ integers have the  form of the wave functions describing 
FQH abelian states in multilayer systems,  with $z_{i}^{\alpha}$ playing the 
role 
of the complex coordinate of the $i^{\rm th}$ electron in the 
$\alpha^{\rm th}$ 
layer (such wave functions were first discussed by Halperin \cite{Halperin}). 
Motivated by this similarity,   
we consider an analytic continuation of  the solutions for anyons in the LLL,  
allowing the parameters $g_{\alpha\beta}$ to take any nonnegative values.    
For a single layer, the edge excitations 
are generated when the function $\Phi_+$ in (\ref{wf}) belongs to the space of 
symmetric polynomials in $\{z_i\}$ \cite{Wen,Wen-review}, which is relevant to 
the polynomial character of the functions  $\Phi_+$ in  (\ref{wf}). 

We now consider a mapping of the system onto a circle. 
Using the operators $\vp^{\alpha}_l$ from (\ref{phi-expan}), 
we  define the fields on the circle (parametrized with an  
 angular coordinate $\theta$) by  
\be
\chi_{\alpha}(\theta) = \sum_{n_{\alpha}=0}^{\infty} \frac{1}{\sqrt{2\pi}} 
\vp^{\alpha}_{n_{\alpha}} e^{in_{\alpha}\theta} \,.
\label{chi-expan}\ee
The commutation relations (\ref{Fermi}) then imply  
\be
 \left\{ \chi_{\alpha}(\theta),\chi_{\beta}(\theta') \right\} 
     &=&
\left\{ \chi_{\alpha}^{\dagger}(\theta),\chi_{\beta}^{\dagger}(\theta') 
  \right\} 
     =0\,, \nonumber \\
\left\{ \chi_{\alpha}(\theta),\chi_{\beta}^{\dagger}(\theta') \right\} 
    & =& \delta_{\alpha\beta}\frac{1}{2\pi}\sum_{n=0}^{\infty}
        e^{i n (\theta-\theta')}
     \equiv \delta_{\alpha\beta} \delta_{\rm per}^+(\theta -\theta') \,,
\label{CR-chi}\ee
where $\delta_{\rm per}^+(\theta -\theta')$ is the  positive frequency part of 
the periodic $\delta$-function.  

We introduce the Hamiltonian for  the fields on the circle corresponding  
to (\ref{H}) as 
\be
H = \frac12 \omega_c^{\rm eff} \sum_{\alpha} \hat N_{\alpha}+
 a \sum_{\alpha}\int_0^{2\pi} d\theta \chi_{\alpha}^{\dagger}(\theta)
    \left( -i\p_{\theta} + \half\sum_{\beta} \lambda_{\alpha\beta} \hat 
N_{\beta} \right)    \chi_{\alpha}(\theta) \,,                                 
\label{H-chi}\ee
with the representation of the particle numbers as 
\be
\hat N_{\alpha} = \int_0^{2\pi} d\theta \chi_{\alpha}^{\dagger}(\theta) 
\chi_{\alpha}(\theta)\,.
\ee
The equations of motion   
\be
\p_t\chi_{\alpha}(\theta) = i \left[ H , \chi_{\alpha}(\theta) \right]
  = -a\left( \p_{\theta} + i \sum_{\beta}\lambda_{\alpha\beta} \hat N_{\beta} 
                     \right) \chi_{\alpha}(\theta)\,,
\ee
on introducing new fields by 
\be
\psi_{\alpha}(\theta) = e^{i\sum_{\beta} 
\lambda_{\alpha\beta}\hat N_{\beta} \theta} \chi_{\alpha}(\theta),
\label{psi}\ee
take a free form, 
\be
(\p_t + a\p_{\theta}) \psi_{\alpha}(\theta) = 0 \,.
\ee

To derive the commutation relations for the $\psi$ operators, it is convenient 
to use the identity
\be
e^{i\sum_{\beta} \lambda_{\alpha\beta}\hat N_{\beta} } \chi_{\gamma}=
\chi_{\gamma} e^{i\sum_{\beta}\lambda_{\alpha\beta}
(\hat N_{\beta}-\delta_{\beta\gamma})} \,.
\label{iden}\ee  
To prove this, we first note that the operators $\chi_{\alpha}$ and 
$\chi_{\alpha}^{\dagger}$ satisfy 
\be
[\hat N_{\alpha},\chi_{\beta}]=-\delta_{\alpha\beta}\chi_{\beta}\,, \quad
[\hat N_{\alpha},\chi_{\beta}^{\dagger}]=
-\delta_{\alpha\beta}\chi_{\beta}^{\dagger}\,. 
\label{N-com-chi}\ee
The first relation in (\ref{N-com-chi}) written as 
$\hat N_{\alpha}\chi_{\beta}=\chi_{\beta}(\hat N_{\alpha}-
\delta_{\alpha\beta})$, 
is generalized  to 
$\hat N_{\alpha}^k\chi_{\beta}=
\chi_{\beta}(\hat N_{\alpha}-\delta_{\alpha\beta})^k$, with 
$k$ a positive integer, which straightforwardly leads to (\ref{iden}). 

Using the identity (\ref{iden}) and similar identities for the operators 
$\chi_{\alpha}^{\dagger}$, we obtain from (\ref{CR-chi}) the commutation 
relations
for the $\psi$ fields:   
 \be
\psi_{\alpha}(\theta)\psi_{\beta }(\theta') + 
e^{-i\lambda_{\alpha\beta}(\theta-\theta')}
 \psi_{\beta }(\theta')\psi_{\alpha}(\theta) &=& 0 \label{comm1} \nonumber\\
\psi_{\alpha}^{\dagger}(\theta)\psi_{\beta}^{\dagger}(\theta') + 
e^{-i\lambda_{\alpha\beta}(\theta-\theta')}
\psi_{\beta}^{\dagger}(\theta')\psi_{\alpha}^ {\dagger}(\theta) 
&=&0\label{CR-psi}\\
\psi_{\alpha}(\theta)\psi_{\beta}^{\dagger}(\theta') + 
e^{i\lambda_{\alpha\beta}(\theta-\theta')}                                
   \psi_{\beta}^{\dagger}(\theta')\psi_{\alpha}(\theta) 
 &=& \delta_{\alpha\beta}\Delta_{\alpha}(\theta - \theta')\,, \nonumber
\ee
where the operators 
\be
\Delta_{\alpha}(\theta - \theta') = 
e^{i\sum_{\beta}\lambda_{\alpha\beta} \hat N_{\beta}(\theta-\theta')} 
   \delta_{\rm per}^+(\theta - \theta')
\ee
behave to some extent like $\delta$-functions on the circle:
\be
\int_0^{2\pi}d\theta'\psi^{\dagger}_{\alpha}(\theta')\Delta_{\alpha}
(\theta' - \theta) =\psi^{\dagger}_{\alpha}(\theta )\,, \quad 
\int_0^{2\pi}d\theta'\Delta_{\alpha}(\theta - \theta')\psi_{\alpha}(\theta' ) =
\psi_{\alpha}(\theta)\, .
\label{delta}\ee

Consider the ``Fourier transform'' of the $\psi$ operators: 
\be
\psi_{\alpha}(\theta) = \frac{1}{\sqrt{2\pi}} 
                 \sum_{\kappa_{\alpha}} a^{\alpha}_{\kappa_{\alpha}} 
e^{i\kappa_{\alpha}\theta}\,\,.
\label{psi-F}\ee
The (nonnegative) numbers $\kappa_{\alpha}$ are not integers, in contrast 
to the 
numbers $n_{\alpha}$ in the expansion (\ref{chi-expan}). 
Allowed values of $\kappa_{\alpha}$ can be obtained by 
calculating the matrix elements of the field $\psi_{\alpha}$ 
between states with fixed particle numbers $\{n'_{\beta}\}$.
The only nonzero matrix elements   
correspond to the transition 
$n'_{\alpha}+1\to n'_{\alpha}$, without changing the other numbers of 
particles.  
Comparing the matrix elements of $\psi_{\alpha}$, in its two representations,
(\ref{psi-F}) and (\ref{psi}) with (\ref{chi-expan}), 
yields the restrictions 
 \be
\kappa_{\alpha} = n_{\alpha} + \sum_{\beta} \lambda_{\alpha\beta} 
                                          n'_{\beta} \,,  
 \label{kappa}\ee
where the nonnegative integers $n_{\alpha}$ are the same as in 
(\ref{chi-expan}).

Let us introduce a  set of basis states
and label  for a moment all the particles by a single  sequence $1, \dots, N$ 
\be
|\kappa_N^{\alpha_N},\cdots, \kappa_1^{\alpha_1} \rangle =\frac{1}{\sqrt{N!}}
a^{\alpha_N\dagger}_{\kappa_N}\cdots a^{\alpha_1\dagger}_{\kappa_1}   
    |0\rangle \,.
\label{basis}\ee 
Then we obtain the restriction 
\be
\kappa^i_{\alpha}=n^i_{\alpha}+\sum_{\beta}\lambda_{\alpha\beta} 
          N_{\beta}^{[i]}\,,
\quad i=1,2,\dots ,N_{\alpha} \;,
\label{kappa=}\ee
where for each species $\alpha$,  $i$ numbers particles in  order of their 
appearance in (\ref{basis}), $n^i_{\alpha}$
correspond to the fermionic angular momentum quantum numbers 
(see (\ref{L(N)})).
We have also denoted by 
$N_{\beta}^{[i]}$ the number of creation operators  for species $\beta$ 
in the sequence (\ref{basis}) (counted from the right to the left)
standing before the operator creating the  
$i^{\rm th}$ particle of species $\alpha$
(in other words, $N_{\beta}^{[i]}$ is the number of particles of 
species $\beta$ created before the 
$i^{\rm th}$ particle of species $\alpha$). 

Inserting the expansion (\ref{psi-F}) into (\ref{CR-psi}), we obtain the 
commutation relations for the operators 
$a_{\kappa}^{\alpha}$ and $a_{\kappa}^{\alpha \dagger}$ 
(here we drop the index of $\kappa$ labelling 
the species of particles) 
\be
a_{\kappa}^{\alpha} a_{\mu}^{\beta } 
+ a_{\mu - \lambda_{\alpha\beta}}^{\beta } 
a_{\kappa + \lambda_{\alpha\beta}}^{\alpha}
&=&0\,, \nonumber\\
a_{\kappa}^{\alpha \dagger} a_{\mu}^{\beta \dagger} 
+ a_{\mu + \lambda_{\alpha\beta}}^{\beta \dagger} a_{\kappa - 
\lambda_{\alpha\beta}}^{\alpha \dagger}
&=&0 \,,\label{ocom} \\
a_{\kappa}^{\alpha} a_{\mu}^{\beta \dagger} 
+ a_{\mu - \lambda_{\alpha\beta}}^{\beta \dagger} a_{\kappa - 
\lambda_{\alpha\beta}}^{\alpha}
&=&\delta_{\alpha\beta} \delta_{\kappa\mu} \Pi_{\kappa}^{\alpha}\,\,. \nonumber
\ee
where $\Pi_{\kappa}^{\alpha}$ are the projections  on the subspace with  the
particle numbers $\{ n'_{\beta} \}$ determined by  (\ref{kappa}). 
The projection operators satisfy the relations 
\be
a_{\kappa}^{\alpha\dagger}\Pi_{\mu}^{\beta}=
\Pi_{\mu+\lambda_{\alpha\beta}}^{\beta}a_{\kappa}^{\alpha\dagger}\,\,, \quad
a_{\kappa}^{\alpha}\Pi_{\mu}^{\beta}=
\Pi_{\mu-\lambda_{\alpha\beta}}^{\beta}a_{\kappa}^{\alpha}\,\,.
\label{proj-a}\ee
For rational $\lambda_{\alpha\beta}$ with common denominator $q$, explicit 
expressions for these operators can be derived:
\be
\Pi_{\kappa}^{\alpha} = 
\frac{1}{2\pi}\int_{0}^{2\pi}d\theta \sum_{l=0}^{\infty}
e^{-iq(\kappa-l-\sum_{\beta}\lambda_{\alpha\beta}\hat N_{\beta})\theta}=
\frac{1}{q} \sum_{n=1}^q 
    e^{2\pi in(\kappa - \sum_{\beta}\lambda_{\alpha\beta}\hat N_{\beta})}\,\,.
\label{proj=}\ee
It is easily  seen from (\ref{proj=})  
that on the states with fixed particle numbers, the 
operators (\ref{proj=}) have eigenvalues 1 if the relations (\ref{kappa}) 
are fulfilled and eigenvalues 0 otherwise as it should be. 

The occupation number description 
determined by the  creation and annihilation operators 
$a_{\kappa}^{\alpha\dagger}$ and $a_{\kappa}^{\alpha}$ 
is not unique:
as can be seen from (\ref{ocom}), the 
interchange of any two of the creation operators leads to an
equivalent state corresponding to the same fermionic quantum numbers 
$\{n_{\alpha}^{i}\}$. Rearranging an $N$-particle state
does not change the total energy of the state, which is given by the
sum over single-particle energies,
\be
E = \sum_{\alpha}\sum_{j=1}^{N_{\alpha}} (\texthalf\omega_c^{\rm eff} +  
a {\kappa}^{j}_{\alpha})\,  .
\label{energy-nonint}\ee

An ordering procedure  may be introduced by demanding that each particle
added into the system has to have a higher (or equal) value of $\kappa$ 
than the largest $\kappa$ already present in the system.
This means that only the states (\ref{basis})  
corresponding  to the ordering   $\kappa_1 \leq \cdots\leq \kappa_N$ 
are regarded as physical states, which makes the 
occupation number picture unique. 
Any of such normal ordered states can also be constructed in a unique way 
starting from a given set of the fermionic angular momentum quantum numbers
$\{n_{\alpha}^i\}$
by putting particles into the system one by one and always making sure
to put in the particle which gets the lowest possible value of $\kappa$. 

Eq.~(\ref{kappa=}) then admits  the interpretation that the spectrum
of available single particle energy levels for a particle of a given species,
say $\alpha$, is shifted upwards by the presence of all particles below it. 
Each
particle, say of species $\beta$, causes a shift $\lambda_{\alpha\beta}$.
By  convention, particles of distinct species in the same
level shift each other
by $\frac12 \lambda_{\alpha\beta}$($=\frac12 \lambda_{\beta\alpha}$). 
This means that the order of filling of levels of the same 
energy with particles of distinct species is unessential. 

Figure 1 illustrates the above normal ordering procedure in the case 
of a simple two-component system with integer statistics parameters.

\bigskip
\noindent
{\bf 3. Relation to exclusion statistics}
\medskip

\noindent
{\em 3.1. Thermodynamic limit of the free particle formulation}
\medskip

With the ordering procedure described at the end of the previous section, 
Eq.~(\ref{kappa=}) can be written in a compact form as 
\be
{\kappa}_{j}^{\alpha} = n_{j}^{\alpha} + \sum_{\beta}\sum_{l}
\lambda_{\alpha\beta} h({\kappa}_{j}^{\alpha} - {\kappa}_{l}^{\beta}), 
\label{renorm}\ee
where   $h(x)$ is   the step function, which  is defined here as $h(x)=1$ for 
$x> 0$, $h(x)=\frac12$ for $x=0$, 
 and $h(x)=0$ for $x<0$.  We assume the ordering 
${\kappa}_{1}^{\alpha} < {\kappa}_{2}^{\alpha}<\cdots 
<{\kappa}_{N_{\alpha}}^{\alpha}$ within each species (the resulting numbers 
${\kappa}_j^{\alpha}$ within the same species are distinct). In the form 
(\ref{renorm}), the numbers ${\kappa}_{j}^{\alpha}$ are similar to   
``renormalized'' quantum numbers introduced for integrable models of a   
Calogero type 
\cite{Kawakami-renorm}.  

We now specify the length of the circle as $L$ and use 
the identification \cite{HLV} 
\be
a = \frac{2\pi}{L} v \,.    
\label{fermihast}\ee
If one assumes that the  harmonic  
potential is created in an anyon droplet of radius $L/2\pi$ by an electric 
field, 
then the velocity $v$ can be   interpreted as the drift velocity on the edge 
$E/B$ (the velocity of the edge excitations) where $E=(m/e)\omega^2 R$ is the 
electric field on the edge.  The thermodynamic limit is understood as 
$\omega\to 0$ and $L\to\infty$ while $v$ is kept fixed.
We also introduce the momenta and pseudomomenta
\be
p_j^{\alpha}=\frac{2\pi}{L} n_j^{\alpha},  \quad 
k_j^{\alpha}=\frac{2\pi}{L} {\kappa}_j^{\alpha}
\label{p,k}\ee 
distributed in the thermodynamic limit with
the densities $\nu_{\alpha}(p^{\alpha})$ and 
$\rho_{\alpha}(k^{\alpha})$, respectively, so that 
the number of particles of species $\alpha$ in the interval 
$(p^{\alpha},p^{\alpha}+dp^{\alpha})$ or in the 
corresponding interval $(k^{\alpha},k^{\alpha}+dk^{\alpha})$ 
is\footnote{The transition to the thermodynamic limit below is similar 
to that given in Ref.~\cite{KatoKuramoto}. We stress however that our starting 
point is the free field occupation number picture rather than the spectrum of 
an integrable model as in the above reference.}  
\be
\nu_{\alpha}(p^{\alpha})dp^{\alpha} = \rho_{\alpha}(k^{\alpha})d k^{\alpha}
\label{nu-rho}\ee
In terms of (\ref{p,k}), Eq.~(\ref{renorm}) reads 
\be
p_j^{\alpha}=k_j^{\alpha}-\frac{2\pi}{L}\sum_{\beta}\sum_{l}
 \lambda_{\alpha\beta} 
 h(k_j^{\alpha} - k_l^{\beta}), 
\ee
or, in the thermodynamic limit,
\be
p^{\alpha} = k^{\alpha} - \sum_{\beta}\int_0^{\infty}\lambda_{\alpha\beta} 
 h(k^{\alpha}-k^{\beta})\rho_{\beta}(k^{\beta}) dk^{\beta}.
\label{p-k}\ee
From this equation, we get 
\be
\p p^{\alpha}/\p k^{\alpha} =1-\sum_{\beta}
\lambda_{\alpha\beta} \rho_{\beta}(k^{\alpha})
                                                \label{dp/dk}\ee
and with the use of (\ref{nu-rho}),
\be
\nu^{\alpha} = \frac{\rho^{\alpha}}{1-\sum_{\beta}\lambda_{\alpha\beta}
\rho_{\beta}}.
\ee
The fermionic description of the energy levels  implies that the 
(non-equilibrium) entropy has the form  
\be
S = \frac{L}{2\pi}\sum_{a}\int_0^{\infty}  [- \nu_{\alpha}\ln\nu_{\alpha} - 
                    (1-\nu_{\alpha})\ln(1-\nu_{\alpha}) ] dp^{\alpha} 
\ee
or, in terms of $\rho_{\alpha}$,
\be
S &=& \frac{L}{2\pi}\sum_{\alpha}\int_0^{\infty} \left\{ -\rho_{\alpha}
\ln\rho_{\alpha}
+\left[1-{\textstyle\sum_{\beta}}\lambda_{\alpha\beta}\rho_{\beta}\right]
\ln \left[1-{\textstyle\sum_{\beta}}\lambda_{\alpha\beta}\rho_{\beta}\right]
\right.\nonumber\\ 
&& - \left.  \left[1-{\textstyle\sum_{\beta}}(\delta_{\alpha\beta}+
\lambda_{\alpha\beta})\rho_{\beta}\right]
\ln \left[1-{\textstyle\sum_{\beta}}(\delta_{\alpha\beta}
+\lambda_{\alpha\beta})
\rho_{\beta}\right] \right\}
 dk^{\alpha} \,.
\label{S-excl}\ee
The latter expression, along with the expression for the energy
\be
E = \frac{L}{2\pi} 
\sum_{\alpha}\int_0^{\infty} ({\textstyle \frac12}\omega_c 
+ vk^{\alpha}) \rho_{\alpha}(k^{\alpha}) dk^{\alpha}.
\label{E-excl}\ee
shows that in the thermodynamic limit the model on the circle 
is equivalent to a gas of free chiral particles 
with linear dispersion (with  a gap), the same for all the species,  obeying 
{\it ideal} fractional exclusion statistics  \cite{I-MPLB,Wu} 
with the  fermionic statistics parameters $\lambda_{\alpha\beta}$.  
We refer to the {\it bosonic} and {\it fermionic} statistics parameters 
\cite{I-PRB} $g_{\alpha\beta}$ and $\lambda_{\alpha\beta}$  
as those appearing in the bosonic and fermionic counting of the statistical 
weight  
\be
W=\prod_{\alpha} {(D^b_{\alpha}+N_{\alpha}-1)! \over
N_{\alpha}!(D^b_{\alpha}-1)!}=\prod_{\alpha}{D^f_{\alpha}!
\over N_{\alpha}!(D^f_{\alpha}-N_{\alpha})!}\,.
\label{W}\ee
with  
\be
D_{\alpha}^{b}=G_{\alpha}-\sum_{\beta}g_{\alpha\beta}N_{\beta} , \quad 
D_{\alpha}^{f}=G_{\alpha}-\sum_{\beta}\lambda_{\alpha\beta}N_{\beta} , \quad 
\label{D}\ee
resulting in the expression for the entropy (\ref{S-excl})
upon taking the thermodynamic limit 
$N_{\alpha}\to\infty$, $G_{\alpha}\to\infty$, 
with $\rho_{\alpha}=N_{\alpha}/G_{\alpha}$ kept constant. Eq.~(\ref{W}) 
implies 
the relation between the bosonic and fermionic statistics parameters   
\be
g_{\alpha\beta}=\delta_{\alpha\beta}+\lambda_{\alpha\beta}\;.
\label{g}\ee

Equations for the distribution functions $\rho_{\alpha}$ in the equilibrium 
can be 
derived from Eqs.~(\ref{S-excl},\ref{E-excl}) in the usual way. In terms of 
the  
single state grand partition functions $\xi_{\alpha}$, related to the  
$\rho_{\alpha}$'s by \cite{IMO} 
\be
\xi_{\alpha}=1+\frac{\rho_{\alpha}}{1-\sum_{\beta}(\delta_{\alpha\beta}+
\lambda_{\alpha\beta})\rho_{\beta}}\;,
\label{xi-rho}\ee
these read
\be
(\xi_{\alpha}-1)\prod_{\beta} \xi_{\beta}^{\lambda_{\alpha\beta}}=
e^{(\mu_{\alpha}-\frac12\omega_c-vk^{\alpha} )/T}\;.
\label{xi-alpha}\ee

\bigskip
\noindent
{\em 3.2. Heat capacity}
\medskip

We now assume that all the species are kept at the same chemical potential, 
$\mu_{\alpha}=\mu$.  
We also restrict to  the case 
where all the distributions $\rho_{\alpha}$ (and, correspondingly, 
$\xi_{\alpha}$)
are equal to each other. It then follows from (\ref{xi-alpha}) that  
\be
\lambda M \equiv   \sum_{\beta=1}^M\lambda_{\alpha\beta}
\label{lambda}\ee
is independent of $\alpha$, and 
(\ref{xi-alpha}) coincides with the  equation 
for the partition function for a single species  of particles with     
the statistics parameter (\ref{lambda}):
\be
(\xi_{\alpha}-1)\xi_{\alpha}^{\lambda M}=
e^{(\mu_{\alpha}-\frac12\omega_c-vk )/T} \,.
\label{xi-alpha-1species}\ee

We calculate the low temperature heat capacity for this case following 
essentially Ref.~\cite{IAMP} 
for the case of a constant density of states, which holds for a linear 
dispersion 
in one dimension  
(the difference from  Ref.~\cite{IAMP} is that in the case at hand the 
particles 
are chiral and have a gap in the dispersion law). 
Introducing the variable $w$ by  
 \be
\xi_{\alpha}=\frac{1}{1-w}
\label{w}\ee
we represent the particle density  
$d=N/L=\sum_{\alpha}\int_0^{\infty} \rho_{\alpha}(k) \frac{dk }{2\pi}$
 and the energy (\ref{E-excl}) as integrals  over the variable $w$:
\be
d&=&\frac{MT}{2\pi v } \int_0^{w(0)} \frac{dw}{1-w} , \nonumber\\ 
\frac{E}{L} &=& \frac{MT}{2\pi v } \int_0^{w(0)} \frac{dw}{1-w} 
\left[\mu_{\alpha}+(1+\lambda M)T\ln (1-w) -T\ln w \right] \,
\label{N-E-w}\ee
where $w(0)$ is the solution of Eqs.~(\ref{xi-alpha-1species})  and (\ref{w}) 
for $k=0$. 
From the first equation in (\ref{N-E-w}) we find  
\be
w(0)=1-e^{-2\pi v d/MT}.
\label{w0}\ee
From Eqs.~(\ref{xi-alpha-1species}), (\ref{w})  
we then have 
\be
\mu_{\alpha}={\textstyle \frac12}\omega_c+ 2\pi (\lambda M +1)v \frac{d}{M} 
+ T \ln \left(1-e^{-2\pi  v d/MT}  \right).
\label{mu}\ee
Finally, combining this with the second equation in (\ref{N-E-w}), we obtain 
\be
\frac{E}{L}= d\left[ {\textstyle \frac12}\omega_c+  
(\lambda M +1)\pi v \frac{d}{M}
\right]+
Td\ln w(0)-\frac{MT^2}{2\pi v}
\int_0^{w(0)} dw  \frac{\ln w}{1-w}.
\label{E}\ee
At low temperatures $T\ll 2\pi v d/M$, up to  non-perturbative corrections, 
containing the exponential factor $e^{-2\pi v d/MT}$, 
$w(0)$ can be set to one. The integral in (\ref{E}) then equals 
$-\frac16\pi^2$, which results in the heat capacity
\be
\frac{C_V}{L}=M\frac{\pi }{6 v} T .
\label{c}
\ee
In the single species case this result reduces to that obtained in 
Ref.\cite{HLV} in a different way. The present derivation 
shows that (\ref{c})
exhausts all the perturbative (behaving as powers of the temperature) terms 
in the heat capacity. 

\bigskip
\noindent
{\em 3.3. Asymptotic Bethe ansatz equations }
\medskip

The connection with ideal exclusions statistics suggests that the equations 
of the model can be represented  in a  form similar to  asymptotic  
Bethe ansatz 
equations \cite{Sutherland} (the latter have the form of the    
thermodynamic Bethe ansatz  equations \cite{YY}, with two-body scattering 
phases (matrices) having a step-wise form). 
Indeed,  introducing  the ``dressed'' energies  $\e_{\alpha}$  
by 
\be
\nu_{\alpha}=\frac{1}{e^{\e_{\alpha}/T}+1}
\label{nu-alpha}\ee
so that 
\be
\xi_{\alpha}=1+e^{-\e_{\alpha}/T} \;, 
\label{xi-e}
\ee
we find that Eqs.~(\ref{xi-alpha}) become    
\be
{\e_{\alpha}}=-\mu_{\alpha}+\e^0_{\alpha}(k_{\alpha})+T\sum_{\beta}
\lambda_{\alpha\beta} \, 
\ln \left(1+e^{(\mu_{\beta}-{\e_{\beta}})/T}\right) \;,  
\label{TBA-like}\ee
where 
\be
\e^0_{\alpha}(k_{\alpha})={\textstyle \frac12}\omega_c+vk^{\alpha} \;. 
\label{bareenergy}\ee

Equations (\ref{TBA-like}), along with (\ref{p-k}), have the form of 
the asymptotic Bethe ansatz equations 
for a system of particles with the  ``bare'' energy (\ref{bareenergy})
and  the step-wise ``two-body scattering phases'' 
\be
\theta(k_j^{\alpha}-k_l^{\beta})=
\lambda_{\alpha\beta} h(k_j^{\alpha}-k_l^{\beta})\;. 
\label{phases}\ee
One can in addition introduce the density of holes 
$\rho_{\alpha}^h(k)$, related to the interval $dk^{\alpha}$, by 
\be
(1-\nu_{\alpha}(p^{\alpha}))d p^{\alpha}
=\rho_{\alpha}^h(k^{\alpha})d k^{\alpha}\;.
\label{rho-h}
\ee
With the use of (\ref{nu-rho}) and (\ref{dp/dk}), we obtain
\be
 \rho_{\alpha}^h 
=1-\sum_{\beta}(\lambda_{\alpha\beta}+\delta_{\alpha\beta})\rho_{\beta}\;. 
\label{rho-h=}\ee
For $T=0$,  
in the case of equal boundary pseudomomenta, $k_0^{\alpha}=k_0$, we have 
\be
\begin{array}{lll}
\rho_{\alpha}=\sum_{\beta}(\lambda_{\alpha\beta}+\delta_{\alpha\beta})^{-1}, 
  & \quad  \rho_{\alpha}^h=0; & \quad k< k_0\;; \\
\rho_{\alpha}=0, & \quad \rho_{\alpha}^h=1; &\quad  k>k_0 \;.\\
\end{array}
\label{T=0distributions}\ee

We now turn again to the above special case of equal 
$\rho_{\alpha}$ for all species 
and  evaluate the  Fermi velocities    
of  quasiparticles  (the same for all species)  
in the picture of interacting fermions for $T= 0$ 
\be
v_F^{\alpha}\equiv \left. \frac{\p {\e_{\alpha}}}{\p p}\right|_{p=p_F}=
\left.\left( \frac{\p k}{\p p} 
\frac{\p {\e_{\alpha}}}{\p k}\right)\right|_{k=k_0} \;, 
\label{vF}\ee
where $p_F$ is the Fermi momentum for the distributions $\nu_{\alpha}$ and 
$k_0$ is the boundary (pseudo Fermi) momentum for the 
distributions $\rho_{\alpha}$. 

For zero temperature, 
because of the jumps in the distributions of pseudoparticles and  holes, 
the expression (\ref{vF}) is not well 
defined for $k=k_0$. It is well defined however for $k= k_0^-(\equiv k_0-0)$ 
and 
for $k= k_0^+(\equiv k_0+0)$. 
In the above special case of equal 
distribution functions for all the species, 
we obtain from (\ref{TBA-like}), 
$\e'_{\alpha}(k_0^-)=v(\lambda M +1)^{-1}$ and 
$\e'_{\alpha}(k_0^+)=v$. 
With ${\p p}/{\p k}= (\lambda M+1)^{-1}$ for $k=k_0^-$ and
 ${\p p}/{\p k}= v$ for $k=k_0^+$  
following from  (\ref{dp/dk}),   
we conclude that
\be
v_F^{\alpha}(k_0^-)=v_F^{\alpha}(k_0^+)=v, 
\quad\quad {\rm for }\quad\quad T=0\;.
\ee 
Representing then the Fermi velocities (\ref{vF}) as   
\be
v_F^{\alpha}=\left\{ 
  \begin{array}{ll}
   \e'_{\alpha}(k)/\rho(k), & \quad k=k_0^- ,\\
   \e'_{\alpha}(k), & \quad k=k_0^+ ,\\
  \end{array}
\right.
\label{vF:T=0}
\ee
we remark that the first line in (\ref{vF:T=0}), 
which is normally used to evaluate the Fermi velocities 
in the thermodynamic Bethe ansatz  method (see e. g. \cite{IKR}), 
in the case of step-wise two body phase shifts  (asymptotic Bethe ansatz) 
holds only for $k<k_0$. 

Similar peculiarities due to a 
step-wise form of the two-body scattering phase 
also arise for other quantities determined by the asymptotic Bethe ansatz 
equations. We discuss here the dressed charge matrix (which is   
related to Friedel oscillations, conductivity etc., see \cite{Voit}).  
With the two-body scattering phases (\ref{phases}), 
the dressed charge matrix is determined by the equations \cite{IKR}  
\be
Z_{\alpha\beta}(k_{\beta})=\delta_{\alpha\beta}-\sum_{\gamma}
\int_0^{k_0} Z_{\alpha\gamma}(k'_{\gamma})\, \lambda_{\gamma\beta}
\,\delta (k'_{\gamma}-k_{\beta})\, dk'_{\gamma}.
\label{dressedcharge}\ee 
The function $Z_{\alpha\beta}(k_{\beta})$ thus has a jump at the boundary 
pseudomomenta $k_0$: below the pseudo Fermi level  
\be
Z_{\alpha\beta}(k_0^-)=(g^{-1})_{\alpha\beta}\,,
\label{Z-in}\ee
while above the pseudo Fermi level 
\be
Z_{\alpha\beta}(k_0^+)=\delta_{\alpha\beta} \,.
\label{Z-out}\ee
The relation between the  dressed charge matrix and exclusion 
statistics matrix 
can also be written in a matrix form as    
\be
Z(k_0^-)Z(k_0^+)
=g^{-1}\,.
\label{Z-g}\ee
which is different form the
result $Z(k_0)Z(k_0)=g^{-1}$
 obtained  in Ref.~\cite{FK-JPhysA}. 

Note that the 
representation of the Fermi velocity (\ref{vF:T=0}) and the relation of 
the dressed 
charge matrix to the statistics matrix (\ref{Z-g}) are valid for an  
arbitrary  
``bare'' energy of particles $\e^0_{\alpha}(k)$   
(which in the case at hand  is (\ref{bareenergy})).  
For the case of a single species, the relation similar to (\ref{Z-g}),   
between the dressed charge function and the fractional exclusion statistics 
parameter, was discussed  in Ref.~\cite{WuYu}.

\bigskip
\noindent
{\em 3.4. Thermal excitations}
\medskip

Because of the nontrivial temperature dependence of the dressed energy 
$\e(k)$ 
determined by Eq.(\ref{TBA-like}), it is not obvious that 
the above zero temperature quasiparticle excitations survive 
at finite temperatures. In this subsection,  
using the exclusion statistics  representation of the model, we calculate 
   the Fermi velocities (\ref{vF}) at low but nonzero 
temperatures. The derivatives with respect to $k$ in (\ref{vF}) 
then have  to be evaluated  
at the point $\mu_{\alpha}=\e^0_{\alpha}(k) $. 
 
The derivative ${\p p}/{\p k} $ is given by  (\ref{dp/dk}), 
which in the case at hand reads 
\be
\frac{\p p}{\p k}=1-\lambda M\rho_{\alpha} \;.
\label{dp/dk=}\ee
To calculate the derivative ${\p {\e_{\alpha}}}/{\p k}$, we note that, 
according to (\ref{xi-alpha}) and (\ref{xi-e}), 
$\e_{\alpha}$ depend on $k$ only via 
$x\equiv e^{(\mu_{\alpha}-\e^0_{\alpha}(k) )/T}$, which yields  
\be
\frac{\p {\e_{\alpha}}}{\p k}
=\frac{\p x}{\p k}\frac{\p {\e_{\alpha}}}{\p x}=
vx \frac{\p \xi_{\alpha}}{\p x} \frac{1}{\xi_{\alpha}-1}\;.
\label{de/dk}\ee
Taking into account the relation between the single state 
distribution functions 
and partition functions ($\xi=\prod_{\alpha}\xi_{\alpha}$),
\be
\rho_{\alpha}=x\frac{1}{\xi}\frac{\p \xi}{\p x}= 
M x\frac{1}{\xi_{\alpha}}\frac{\p \xi_{\alpha}}{\p x} \;,
\label{rho-xi}\ee 
and expressing $\xi_{\alpha}$ in terms of $\rho_{\alpha}$ with the use 
of (\ref{xi-rho}), we get 
\be
\frac{\p {\e_{\alpha}}}{\p k}=v(1-\lambda M\rho_{\alpha}) \;.
\label{de/dk=}\ee
Note that both the derivatives (\ref{dp/dk=}) 
and (\ref{de/dk=}) are well defined 
at the point  $\mu_{\alpha}=\e^0_{\alpha}(k) $.  
Using  these, we finally obtain 
the Fermi velocities of quasiparticles for $T\neq 0$
\be
v_{F}^{\alpha}=v\;.
\label{v_a}\ee
This calculation proves that the Fermi velocities  of the  
quasiparticles at nonzero temperatures remain the same as at $T=0$. 
As the derivation shows, the conclusion holds for an arbitrary bare energy 
$\e^0_{\alpha}(k) $ in the asymptotic Bethe ansatz equations (\ref{TBA-like}). 

\bigskip
\noindent
{\bf 4. Bosonization and relation to edge excitations in multilayer FHQ states}
\medskip

In this section we  bosonize the model governed by the Hamiltonian 
(\ref{H-chi})  
generalizing the procedure given in Ref.~\cite{HLV} to  the case of 
several chiral fields. We then discuss the relation of the bosonized form 
of the 
model to the effective low energy description  
of edge excitations in multilayer FQH states \cite{WenZee,Wen-review}. 
  
Consider the Hamiltonian (\ref{H-chi}) of chiral fields on the circle of  
length $L$ specified by (\ref{fermihast}). 
The fact that the model is described with only one velocity $v$ means that 
in the picture of the anyon droplet, all the species of particles have 
the same 
boundary radius.   
We assume  that the boundary (pseudo Fermi) values of $\kappa$ 
in the ground state are the same for all species. It then follows from the 
occupation 
rules and the ordering procedure described in Section 2 (or, equivalently, 
from 
Eq.~(\ref{renorm})) that the pseudo Fermi angular momentum is 
\be
\kappa_0^{\alpha} = (\lambda_{\alpha\alpha}+1)(N_{0\alpha} - 1) 
           + \sum_{\beta(\neq \alpha)}\lambda_{\alpha\beta}N_{0\beta},
\ee
where  $N_{0\alpha}$ is the number of particles of species $\alpha$ in 
the ground 
state. 
We introduce the  chemical potentials for zero temperature by 
\be
\mu_0^{\alpha} = \half\omega_c^{\rm eff} + a\kappa_F^{\alpha}\;,
\ee 
with 
\be
\kappa_F^{\alpha} = (\lambda_{\alpha\alpha}+1)\left( N_{0\alpha} 
                  - \half \right) 
           + \sum_{\beta(\neq \alpha)}\lambda_{\alpha\beta}N_{0\beta}\;.
\label{kf}\ee
This choice corresponds to the chemical potential lying on the 
midway between the 
highest occupied and the lowest unoccupied energy levels.

The pseudo Fermi levels in the model on the circle  are directly related to the
physical size of the anyon droplet: for  large $N_{\alpha}$ one has  
\be
R^2\simeq r_0^2 \kappa_F^{\alpha}\simeq r_0^2 
\sum_{\beta}g_{\alpha\beta}N_{0\beta}\,,  
\label{radius}\ee
where $r_0=\sqrt{2/eB_{\rm eff}}$ is the effective magnetic length.  
This means that all 
$\kappa_F^{\alpha}$ have the same value, $\kappa_F$, 
from which follows that 
$\sum_{\beta}g_{\alpha\beta}N_{0\beta}$ should not depend on $\alpha$. 
This also implies  that $\mu_0^{\alpha}=\mu_0$. 

We fix the ground state subtracting  $\mu_0 \hat N$ from the 
Hamiltonian (\ref{H-chi}),
\be
H' \equiv H - \mu_0 \sum_{\alpha}\hat N_{\alpha}
   = a\sum_{\alpha} \int_0^{2\pi} d\theta \chi_{\alpha}^{\dagger}(\theta)
    \left[ (-i\p_{\theta} - \kappa_F) + \half \sum_{\beta}\lambda_{\alpha\beta}
\hat N_{\beta} \right]
    \chi_{\alpha}(\theta)\;.            
\label{H1}\ee
By normal ordering the Hamiltonian with respect to the Fermi levels, 
subtracting a constant and  redefining the field operators 
$\chi_{\alpha}$ for the rest of this section by 
$\chi_{\alpha}\to e^{-iN_{0\alpha}\theta}\chi_{\alpha}$, 
we obtain the Hamiltonian describing the excitations  
\be
H' = a\sum_{\alpha}\left[\int_0^{2\pi} d\theta 
   :\chi_{\alpha}^{\dagger}(\theta) \left( -i\p_{\theta} + \half \right)
    \chi_{\alpha}(\theta):
  + \half\sum_{\beta} \lambda_{\alpha\beta} Q_{\alpha} Q_{\beta}\right]\;,
\label{H'}\ee
where  $Q_{\alpha} = \hat N_{\alpha} - N_{0\alpha}$ are the charge operators. 

The fermion fields $\chi_{\alpha}$ admit the representation 
(for details of the bosonization procedure, see Ref.~\cite{HaldaneJPhysC})
\be
\chi_{\alpha}^{\dagger}(\theta) = \frac{e^{\frac{i}{2}\theta}}{\sqrt{2\pi}}
                    ~ e^{i\frac{\pi}{2}\sum_{\beta\neq \alpha}
                     {\rm sgn}(\alpha - \beta)Q_{\beta}}
                           :e^{i\phi_{\alpha}(\theta)}: \;,
\label{chi-bosonized}\ee
where the chiral boson fields 
are given by  
\be
\phi_{\alpha}(\theta) = \phi_0^{\alpha} - Q_{\alpha} \theta +
i\sum_{n>0}\frac{1}{\sqrt{n}}
  \left[ a_{\alpha n}e^{in\theta} 
  - a_{\alpha n}^{\dagger} e^{-in\theta} \right] \;,
\ee
with   $[\phi_0^{\alpha},Q_{\beta}]=i\delta_{\alpha\beta}$
and 
$[a_{\alpha m}, a_{\beta n}^{\dagger}] = \delta_{\alpha\beta}\delta_{mn}$.
The normal ordering in (\ref{chi-bosonized}) refers only to $a_{\alpha n}$
and $a_{\alpha n}^{\dagger}$. 
The fields $\phi_{\alpha}$ 
satisfy the commutation relations 
\be
\left[ \phi_{\alpha}(\theta), \phi_{\beta}(\theta') \right] 
     = i\pi \delta_{\alpha\beta} \, {\rm sgn}_{\rm per}(\theta-\theta')\;, 
\label{CR-phi}\ee
where ${\rm sgn}_{\rm per}(\theta)$ is the periodic sign function: 
$(\partial/\partial \theta)\, {\rm sgn}_{\rm per}(\theta) 
=2\delta_{\rm per}(\theta)$.
The extra phase  in (\ref{chi-bosonized}) containing the charge operators  
ensures that the fermion operators for distinct species  anticommute 
(rather than commute) with each other.

The bosonized form  of the Hamiltonian (\ref{H'})  reads 
\be
H' &=& a \sum_{\alpha}\left[\frac{1}{4\pi}\int_0^{2\pi} 
      d\theta :(\p_{\theta}\phi_{\alpha})^2: + 
\half \sum_{\beta}\lambda_{\alpha\beta} Q_{\alpha} Q_{\beta}\right]\\
  &=& a\sum_{\alpha}\left[ \sum_n na_{\alpha n}^{\dagger}a_{\alpha n} + 
\half \sum_{\beta} g_{\alpha\beta}Q_{\alpha} Q_{\beta}
             \right],
\ee
where $g_{\alpha\beta}$ are  the bosonic statistics parameters 
(see (\ref{g})). 

Since $g_{\alpha\beta}$ is a symmetric matrix, it can be diagonalized by an
orthogonal transformation, 
$(O^{-1}gO)_{\alpha\beta}=\Lambda_{\alpha}\delta_{\alpha\beta}$. 
The Hamiltonian then becomes that of a set of
uncoupled components with (bosonic) statistics parameters $\Lambda_{\alpha}$,
\be
H' = a\sum_{\alpha}\left[ \sum_n n\at_{\alpha n}^{\dagger}\at_{\alpha n} + 
\half\Lambda_{\alpha} \Qt_{\alpha}^2 \right],   
\label{Hdiag}\ee
where $\at_{\alpha n} = O^{-1}_{\alpha\beta}a_{\beta n}$, $\Qt_{\alpha} = 
O^{-1}_{\alpha\beta}Q_{\beta}$. The transformed operators
$\at_{\alpha m}$ and $\at_{\beta n}^{\dagger}$ still obey the 
bosonic commutation
relations. 

At this stage one has to demand that all the eigenvalues 
$\Lambda_{\alpha}$ of the 
matrix $g_{\alpha\beta}$ should be positive, in order for the Hamiltonian 
(\ref{Hdiag}) to be positive definite. This means, as we will see below, that 
all the 
edge modes in our model propagate in the same direction.    
Note that the diagonal form  of the Hamiltonian (\ref{Hdiag}) implies that the 
low-temperature heat capacity is just  a sum of heat capacities from all the 
species,  ${\pi}T/6v $ from each, in agreement with (\ref{c}). 

The Hamiltonian (\ref{Hdiag}) can be transformed to a
free form 
\be
H'=\frac{a}{4\pi}\sum_{\alpha}\int_0^{2\pi}d\theta :(\p_{\theta}
\phit_{\alpha})^2: \;,
\ee
where the new Bose fields are defined by 
\be
\phit_{\alpha}(\theta) = \frac{1}{\sqrt{\Lambda_{\alpha}}}\phit_0^{\alpha}
                - \sqrt{\Lambda_{\alpha}}\Qt_{\alpha}\theta
                + i\sum_{n>0}\frac{1}{\sqrt{n}}
                  \left[ \at_{\alpha n}e^{in\theta} 
                - \at_{\alpha n}^{\dagger} e^{-in\theta} \right]\,,
\ee
with  $[\phit_0^{\alpha},\Qt_{\beta}] = i\delta_{\alpha\beta}$.

One can define the charged operators
corresponding to the fields $\phit_{\alpha}$:
\be
\chit_{\alpha}^{\dagger}(\theta) = 
\frac{e^{\frac{i}{2}\theta\gamma_{\alpha}^2}}
{\sqrt{2\pi}}\,e^{i\frac{\pi}{2}\sum_{\beta\neq \alpha}{\rm sgn}(\alpha - 
\beta)\Qt_{\beta}\sqrt{\Lambda_{\beta}} / {\gamma}_{\beta}}
   :e^{i{\gamma}_{\alpha}\phit_{\alpha}(\theta)}:\;.
\label{charge}\ee
These operators have charges 
$\tilde q_{\alpha} = {\gamma}_{\alpha}/\sqrt{\Lambda_{\alpha}}$
(determined by the commutator 
$[\tilde Q_{\alpha},\tilde \chi_{\alpha}^{\dagger}]$)
as well as  ``statistics'' ${\gamma}_{\alpha}^2$ in the sense that
\be
\chit_{\alpha}^{\dagger}(\theta)\chit_{\alpha}^{\dagger}(\theta')
= e^{i\pi{\gamma}_{\alpha}^2 {\rm sgn}_{\rm per}(\theta-\theta')}
\chit_{\alpha}^{\dagger}(\theta')\chit_{\alpha}^{\dagger}(\theta) \;.
\ee
In particular, if one chooses  ${\gamma}_{\alpha} = \sqrt{\Lambda_{\alpha}}$ 
and 
if  $\Lambda_{\alpha} = 2m_{\alpha}+1$, with $m_{\alpha}$ a positive integer, 
the  operator (\ref{charge}) is an ``electron'' operator of the kind 
discussed by 
Wen \cite{Wen}. Another special case 
${\gamma}_{\alpha} = 1/\sqrt{\Lambda_{\alpha}}
 = 1/\sqrt{2m_{\alpha}+1}$
corresponds to a particle with fractional charge and statistics equal
to $1/(2m_{\alpha}+1)$ like a fundamental quasiparticle in a 
$\nu = 1/(2m_{\alpha}+1)$ FQH layer.
The fields corresponding to distinct species  anticommute with each other,  
$\{\chit_{\alpha }^{\dagger}(\theta) , \chit_{\beta}^{\dagger}(\theta')\} =0$, 
etc.

A general charged operator can be composed of operators of the above type:
\be
\chi^{\dagger}(\theta) =
\frac{e^{\frac{i}{2} \theta \sum_{\alpha} \gamma_{\alpha}^2}}{(2\pi)^{M/2}}
: e^{i\sum_{\alpha}{\gamma}_{\alpha}\phit_{\alpha}(\theta)}: \;\;, 
\label{chop}\ee
which corresponds to the total charge $\sum_{\alpha\beta}O_{\alpha\beta}
\gamma_{\beta}/\sqrt{\Lambda_{\beta}}$ 
associated with  the operator $\sum_{\alpha}Q_{\alpha}$.

To discuss the relation to 
the edge excitations in multilayer FQH systems \cite{WenZee,Wen-review}, 
we  rescale the  Bose fields
$\phit_{\alpha}$ and 
transform back to the non-diagonal picture, using the inverse of
the above orthogonal transformation. This defines new fields
\be
\phib_{\alpha} \equiv O_{\alpha\beta}
\frac{1}{\sqrt{\Lambda_{\beta}}}\phit_{\beta}\;,
\ee
obeying the commutation relations (cf. (\ref{CR-phi}))
\be
\left[ \phib_{\alpha}(\theta), \phib_{\beta}(\theta') \right] 
     = i\pi g^{-1}_{\alpha\beta}\, {\rm sgn}_{\rm per}(\theta-\theta')\;. 
\label{phiWen}\ee
The Hamiltonian (\ref{H'}) takes the form
\be
H' = \frac{a}{4\pi}\sum_{\alpha\beta}g_{\alpha\beta}
  \int_0^{2\pi}d\theta:(\p_{\theta}\phib_{\alpha})
                       (\p_{\theta}\phib_{\beta}):\;\;.
\label{HWen}
\ee

For the case where the diagonal elements   of the  matrix $g_{\alpha\beta}$ 
are odd 
numbers and the off-diagonal elements are integers, Eqs.~(\ref{phiWen}),
(\ref{HWen}) reduce to those describing   
edge excitations in multilayer abelian FQH states \cite{Wen-review} if one 
identifies 
\be
g_{\alpha\beta}=K_{\alpha\beta}\;, \quad vg_{\alpha\beta}=V_{\alpha\beta}\;,
\label{g_K}\ee  
where  $K_{\alpha\beta}$ 
is the topological matrix (in the symmetric basis) and  $V_{\alpha\beta}$ is 
a positive definite matrix describing the interaction between the chiral 
bosonic 
modes \cite{Wen-review}.
Thus, our model with only one
velocity $v$,  corresponds to the special case where the matrices 
$K$ and  $V$ are proportional to each other.
In the more general case of the effective  Luttinger liquid description  
of the edge excitations, when different edge modes  
may have different velocities, 
the matrices $V$ and  $K$ can be simultaneously diagonalized, and 
the velocities  are determined by the ratios of their respective 
eigenvalues \cite{Haldane-preprint}. 

Note that the conclusion about the identification of the exclusion 
statistics matrix with the topological matrix (the first relation in 
(\ref{g_K})) 
was previously  obtained by Fukui and Kawakami  
for hierarchical FQH states (by comparing  the excitation spectrum for the 
edge excitations and chiral particles obeying ideal fractional exclusion 
statistics) \cite{FK-PRB}.

In addition, 
expressing  the  general charged operator (\ref{chop})
in terms of the fields $\phib_{\alpha}$ as
\be
\chi^{\dagger}(\theta) =
\frac{e^{\frac{i}{2} \theta 
\sum_{\alpha\beta} l_{\alpha}g_{\alpha\beta}^{-1}l_{\beta} }}{(2\pi)^{M/2}}
  :e^{i\sum_{\alpha}\l_{\alpha}\phib_{\alpha}(\theta)}:\;,
\ee
with  $l_{\alpha} = O_{\alpha\beta}\sqrt{\Lambda_{\beta}}{\gamma}_{\beta}$,
we find that this operator has the total charge
\be
Q_{\bf l} = \sum_{\alpha\beta} l_{\beta} g^{-1}_{\alpha\beta}
\ee
and statistics  $\sum_{\alpha\beta} l_{\alpha}g_{\alpha\beta}^{-1}l_{\beta}$.
For integer $\l_{\alpha}$, 
this  agrees with the expressions obtained by Wen and Zee in the
symmetric basis \cite{WenZee}. Finally, the filling factor is
\be
\nu = \sum_{\alpha\beta} g^{-1}_{\alpha\beta} 
    = \sum_{\alpha } \tilde t_{\alpha} \frac{1}{\Lambda_{\alpha}} 
                     \tilde t_{\alpha}
\label{fillfact}\ee
where $\tilde t_{\alpha} = \sum_{\beta} O^{-1}_{\alpha\beta}$. 
Figure 1 illustrates 
the first equality in (\ref{fillfact})
for a simple example of  two components.

\bigskip
\noindent
{\bf 5. Concluding Remarks}

\medskip
We have constructed a model describing  several chiral fields on the circle 
which has the same quantum numbers as a system of anyons of several species 
in the lowest Landau level. The parameters of the model range in wider 
intervals 
than the original (statistics) parameters of anyons; this is achieved by the 
analytic continuation of the solutions corresponding to anyons in the LLL. 

The model incorporates some features  of the  
physics described by the electron wave functions corresponding to   
abelian multilayer FQH states. The harmonic potential added into 
the system of anyons plays the role of the confining potential of electrons 
in the FQH states. The model on the circle was found to recover the 
effective low 
energy description 
of edge excitations in the special case of equal velocities of all edge modes. 
In this sense, the model obtained 
can be considered as a possible dynamic theory  underlying the effective 
chiral Luttinger liquid description of edge excitations. 

In this context we note that some  of the   
correlation functions as well as low temperature thermodynamics 
of edge excitations have been 
obtained using only the effective Luttinger liquid description of the edge 
excitations (see e.~g. the review \cite{Wen-review}). 
On the other hand, recent calculations  of 
nonequilibrium transport properties through a point contact in a Luttinger 
liquid (which is  related to tunneling transport between edge states in FQH 
devices) are based on a  particular (integrable) dynamic model of the 
Luttinger liquid \cite{FLS}. 

In this connection the question arises to what extent  these nonequilibrium 
properties depend on 
a  particular dynamic model of edge excitations. 
The model we have discussed  has  a simple dynamics encoded 
in  step-wise two-body scattering phases when it  is formulated in the 
form of   
the asymptotic Bethe ansatz equations (see Sect.~3).  
  It seems therefore to be interesting to use this  model 
to investigate  the above issue. 

Another remark concerns the edge excitations for hierarchical FQH states. 
Integrable models with long range interactions can be constructed 
which are described by the same matrices as the topological matrices 
corresponding to hierarchical FQH states  (see \cite{Kawakami-hier}).  
A simple link between the integrable models with long range  interactions 
and edge excitations for FQH states is given by  the (ideal) fractional 
exclusion statistics. Exclusion statistics provides, in addition, a connection 
with anyons in the LLL \cite{Wu,dVO,IMO}. This suggests that an explicit   
dynamic model of edge excitations for  the hierarchical FQH states  
can also be constructed, along the lines of the present paper,
starting from the appropriate picture of anyons in the LLL.

\vskip5mm
\centerline{\bf Acknowledgements}

\medskip
We are grateful to T.~H.~Hansson and J.~M.~Leinaas
for numerous discussions and suggestions.
We would also like to thank S.~Mashkevich for an important remark. 

\eject

\eject
\phantom{$a$}

\phantom{$a$}

\vspace{2cm}

\setlength{\unitlength}{0.012500in}%
\begin{center}
\begin{picture}(100,320)(240,240)
\thicklines
\put(270,260){\circle{14}}
\put(270,400){\circle{14}}
\put(310,260){\circle*{14}}
\put(310,540){\circle*{14}}
\put(270,540){\circle{14}}
\put(240,280){\line( 1, 0){100}}
\put(240,320){\line( 1, 0){100}}
\put(240,360){\line( 1, 0){100}}
\put(240,400){\line( 1, 0){ 20}}
\put(240,440){\line( 1, 0){100}}
\put(240,520){\line( 1, 0){100}}
\put(240,560){\line( 1, 0){100}}
\put(240,240){\line( 1, 0){100}}
\put(280,400){\line( 1, 0){ 60}}
\put(240,480){\line( 1, 0){100}}
\multiput(240,300)(8.00000,0.00000){13}{\line( 1, 0){  4.000}}
\multiput(240,340)(8.00000,0.00000){13}{\line( 1, 0){  4.000}}
\multiput(240,380)(8.00000,0.00000){13}{\line( 1, 0){  4.000}}
\multiput(240,420)(8.00000,0.00000){13}{\line( 1, 0){  4.000}}
\multiput(240,460)(8.00000,0.00000){13}{\line( 1, 0){  4.000}}
\multiput(240,500)(8.00000,0.00000){13}{\line( 1, 0){  4.000}}
\multiput(240,540)(8.00000,0.00000){3}{\line( 1, 0){  4.000}}
\multiput(280,540)(8.00000,0.00000){3}{\line( 1, 0){  4.000}}
\multiput(320,540)(8.00000,0.00000){3}{\line( 1, 0){  4.000}}
\multiput(240,260)(8.00000,0.00000){3}{\line( 1, 0){  4.000}}
\multiput(280,260)(8.00000,0.00000){3}{\line( 1, 0){  4.000}}
\multiput(320,260)(8.00000,0.00000){3}{\line( 1, 0){  4.000}}
\put(350,238){\makebox(0,0)[lb]{\raisebox{0pt}[0pt][0pt]{\bf 
{\Large $\kappa = 0$}}}}
\put(350,318){\makebox(0,0)[lb]{\raisebox{0pt}[0pt][0pt]{\bf 
{\Large $\kappa = 2$}}}}
\put(350,398){\makebox(0,0)[lb]{\raisebox{0pt}[0pt][0pt]{\bf 
{\Large $\kappa = 4$}}}}
\put(350,478){\makebox(0,0)[lb]{\raisebox{0pt}[0pt][0pt]{\bf 
{\Large $\kappa = 6$}}}}
\put(350,558){\makebox(0,0)[lb]{\raisebox{0pt}[0pt][0pt]{\bf 
{\Large $\kappa = 8$}}}}
\put(290,580){\makebox(0,0)[lb]{\raisebox{0pt}[0pt][0pt]{\bf .}}}
\put(290,600){\makebox(0,0)[lb]{\raisebox{0pt}[0pt][0pt]{\bf .}}}
\put(290,620){\makebox(0,0)[lb]{\raisebox{0pt}[0pt][0pt]{\bf .}}}
\end{picture}
\end{center}

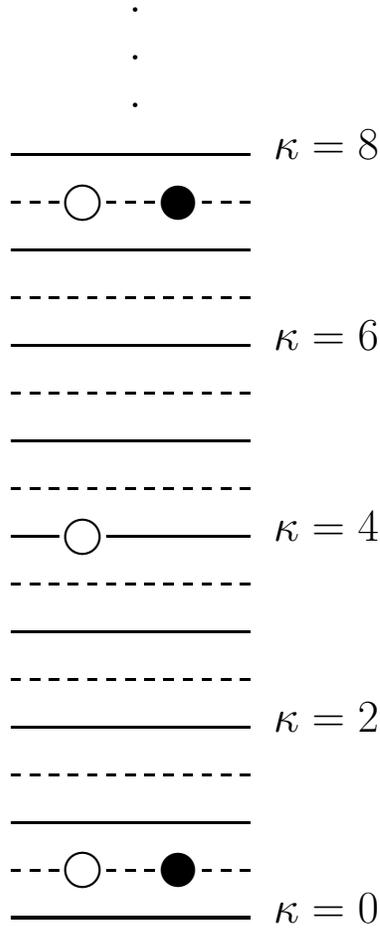
{\bf Fig.~1:} Ground state of a two-component system with 
$g_{11} =3$, $g_{22} = 5$ and $g_{12} = g_{21} = 1$ (bosonic representation). 
The white and black circles denote
component 1 and component 2, respectively.  
The state has been built
up according to the normal ordering convention, with increasing energy
($\kappa$) values. 
The pattern  repeats periodically. 
The filling factor is 
$3/7 = \sum_{\alpha\beta} g^{-1}_{\alpha\beta}$ (if the number of particles
in the ground state is such that the Fermi level is the same for both species).

\end{document}